\documentclass{revtex4}
\usepackage{graphicx}
\usepackage{dcolumn}
\usepackage{amsmath}
\usepackage{array}
\usepackage{bm}
\usepackage{amssymb}
\usepackage{amsfonts}
\begin{document}

\title{Meson spectrum in SU($N$) gauge theories with quarks in higher representations: A check of Casimir scaling hypothesis}

\author{Fabien \surname{Buisseret}$^{1,2}$}
\email[E-mail: ]{fabien.buisseret@umons.ac.be}

\author{Claude \surname{Semay}$^1$}
\email[E-mail: ]{claude.semay@umons.ac.be}
\affiliation{$^1$ Service de Physique Nucl\'{e}aire et Subnucl\'{e}aire,
Universit\'{e} de Mons, UMONS  Research
Institute for Complex Systems, Place du Parc 20, 7000 Mons, Belgium \\
$^2$ CeREF, Chauss\'{e}e de Binche 159, 7000 Mons, Belgium}

\date{\today}

\begin{abstract}
Gauge theories with gauge group $SU(N)$ and quarks belonging to arbitrary representations of $SU(N)$ form a rich landscape of QCD-like theories, whose study can shed new light on the properties of confinement. Four cases are particularly worth of interest: quarks in the fundamental representation, quarks in the 2-indice (anti)symmetric representation and quarks in the adjoint representation. The last three corresponding QCD-like theories are equivalent at large-$N$ for bosonic observables: It is the orientifold equivalence. The behavior of the lightest vector meson mass versus $N$ has been studied in quenched lattice QCD in the chiral limit in these theories. We show that the observed behaviors are compatible with a string tension proportional to the quadratic Casimir of $SU(N)$ in the quark color representation, \textit{i.e.} with the Casimir scaling hypothesis. The large-$N$ limit of some excited meson masses computed in quenched lattice QCD with quarks in the fundamental representation is also shown to be compatible with QCD string's well-known signature: Regge trajectories.
\end{abstract}

\maketitle

\section{Introduction}

The pioneering works of 't Hooft \cite{tHooft:1973alw} and Witten \cite{Witten:1979kh} about the large-$N$ limit of QCD have generated a tremendous amount of studies devoted to the behavior of SU($N$) gauge theories with arbitrary $N$, and when $N\to\infty$ in particular. The interested reader may find recent results in hadronic physics at large-$N$ in the reviews \cite{Manohar:1998xv,Lucini:2012gg}  

A minimal requirement for a SU($N$) gauge theory to mimic QCD at large $N$ is that asymptotic freedom is present for any $N\geq 2$. It has been shown in \cite{Dietrich:2006cm} that only 4 quark color representations $R$ fulfill this requirement when $N_f=2$:
\begin{itemize}
\item Fundamental (fQCD): It is the well-known 't Hooft limit \cite{tHooft:1973alw};
\item Antisymmetric (asQCD): Proposed in \cite{Armoni:2003fb}, it is equivalent to fQCD for $N=3$ but different when $N>3$ (it does not exist for $N=2$). In particular, quark and gluon loops have the same $N$-counting in asQCD;
\item Adjoint (adjQCD): The adjoint representation leads to SUSY QCD if $N_f=1$ \cite{Salam:1975bv};
\item Symmetric (sQCD): It is relevant in the framework of some technicolor models \cite{Sannino:2004qp}.
\end{itemize} 
Note that in the four cases defined above the large-$N$ limit is taken such that 't Hooft coupling $g^2 N$ is constant, $g$ being the strong coupling constant. As $N$ goes to infinity, a particular case of orientifold equivalence is that asQCD, adjQCD and sQCD are equivalent in the bosonic sector \cite{Armoni:2003gp,Armoni:2004ub}. Remarkably, this equivalence tells that asQCD and sQCD are equivalent to SUSY QCD for $N_f=1 $ \cite{Armoni:2003fb}.

Orientifold equivalence has been checked in quenched lattice QCD \cite{Lucini:2010kj} where it is shown that, in the chiral limit, the lightest vector meson masses are compatible when $N\to\infty$ in asQCD, adjQCD and sQCD. More precisely the vector meson masses $M$ have been computed for $N=2$, 3, 4 and 6. So far no model has reproduced the global trends of $M(N)$ observed in the lattice results (displayed in Fig.~\ref{figres}). The main goal of this paper is to show that it is a direct consequence of the Casimir scaling of the string tension \cite{DelDebbio:2001sj,Semay:2004}. 

\section{The model}\label{sec:mes}

At lowest order in the strong coupling expansion of lattice QCD, it can be computed that the string tension of a flux tube starting from a color source in the representation $R$ scales as $\sigma_R =C_R\, g^2 \Omega$ \cite{DelDebbio:2001sj}, where $\Omega$ is some constant and where $C_R$ is the eigenvalue of the quadratic Casimir operator of SU($N$) in the representation $R$, namely $C_R = \bm T_R\cdot\bm T_R$ with $\bm T_R$ the generators of $SU(N)$ in the representation $R$ (bold symbols denote vectors). Introducing the 't Hooft coupling, we can write $\sigma_R=(C_R/N) \sigma_0$ and assume that $\sigma_0$ does not depend on the number of colors: It is the Casimir scaling hypothesis (see \cite{Bali:2000un} for a numerical validation of this hypothesis on static sources). For completeness, we mention that one expects $\sigma_0\approx (9/4)\times(0.17-0.22)$ GeV$^2$ in order to be consistent with SU(3) QCD. It is worth mentioning that the Casimir scaling is also valid at finite temperature, as shown by Polyakov loop computations in full lattice QCD \cite{Petr2015} and in perturbative calculations at next-to-next-to leading order \cite{Berw2016}.

We now reasonably assume that the meson spectrum in the chiral limit, \textit{i.e.} with vanishing bare quark and pion masses, only depends on $\sqrt{\sigma_R}$ at the dominant order. In other words, the confining interaction generates a unique energy scale that sets the scale of the whole spectrum as $M\propto \sqrt{\sigma_R}$. So, we assume that
\begin{equation}\label{model}
M=\sqrt{\frac{C_R}{N}}\, \lambda,
\end{equation}
where $\lambda$ is a parameter to fit. Equation (\ref{model}) is our proposal to model $M(N)$ in any of the QCD-like theories previously defined. The Casimir eigenvalues are given by 
\begin{equation}
C_{\textrm{fQCD}}=\frac{N^2-1}{2N},\ C_{\textrm{asQCD}}=\frac{(N+1)(N-2)}{N},\ C_{\textrm{sQCD}}=\frac{(N+2)(N-1)}{N},\ C_{\textrm{adjQCD}}=N.
\end{equation}

The proposed scaling can be further discussed within the framework of Coulomb gauge QCD, which is a constituent approach of QCD able to reach the chiral limit and to accurately reproduce the light meson spectrum \cite{LlanesEstrada:1999uh,Ligterink:2003hd}. We now summarize the main points of \cite{LlanesEstrada:1999uh} and discuss its generalization to an arbitrary number of colors in the case of the chiral limit ($m_q=0$). 

The quark self-energy in the rainbow approximation allows to find the Bogolioubov angle through the equation
\begin{equation}\label{cg1}
k\, s_k=\int \frac{d^3\bm p}{(2\pi)^3}\hat V(\vert\bm k-\bm p\vert) \left( s_k c_p \hat{ \bm k}\cdot\hat{ \bm p}-s_p c_k\right),
\end{equation} 
where  $c_p$ and $s_p$ are the cosine and sine of the Bogolioubov angle and where $\hat{\bm k}$ is the angular part of the vector $\bm k$. The Fourier transform of linear confining potential with Casimir scaling is used:
\begin{equation}
\hat V(\vert\bm k-\bm p\vert)=-\frac{4\pi\sigma_R}{\vert\bm k-\bm p\vert^4}.
\end{equation}

The gap energy reads 
\begin{equation}\label{cg2}
\epsilon_k=k c_k-\int \frac{d^3p}{(2\pi)^3}\hat V(\vert\bm k-\bm p\vert)\left( c_k c_p \hat{ \bm k}\cdot\hat{ \bm p}+s_p s_k\right).
\end{equation}
Meson masses can finally be computed. In the $1^{--}$ channel, the partial wave equation reads
\begin{equation}\label{cg3}
\left(M-2\epsilon_k\right)\psi(k)=\int^\infty_0\frac{dp\, p^2}{12\pi^2} K_{1^{--}}(k,p)\psi(p),
\end{equation}
with $K_{1^{--}}(k,p)=2c_kc_p\hat V_1+(1+s_k)(1+s_p)\hat V_0$, $\hat V_i$ being the angular integral over $x=\hat{\bm k}\cdot\hat{\bm p}$ of $\hat V(\vert\bm k-\bm p\vert)$ times $x^i$.

It is sufficient for our purpose to notice that equations (\ref{cg1})--(\ref{cg2}) can be rewritten in terms of dimensionless variables through the scaling $\bm p \to \bm p\sqrt{\sigma_R}$ (momentum variables). It then appears that the dynamically generated quark mass $\epsilon_k$ is proportional to $\sqrt{\sigma_R}$ as well as the interaction part of (\ref{cg3}). Hence the meson masses necessarily scale as guessed in (\ref{model}). Note that the same conclusion would be reached by Bethe-Salpeter-based approaches like in \cite{Bicudo:1989sh}. 

\section{Results and conclusion}

\begin{figure}[ht]
\includegraphics[width=11cm]{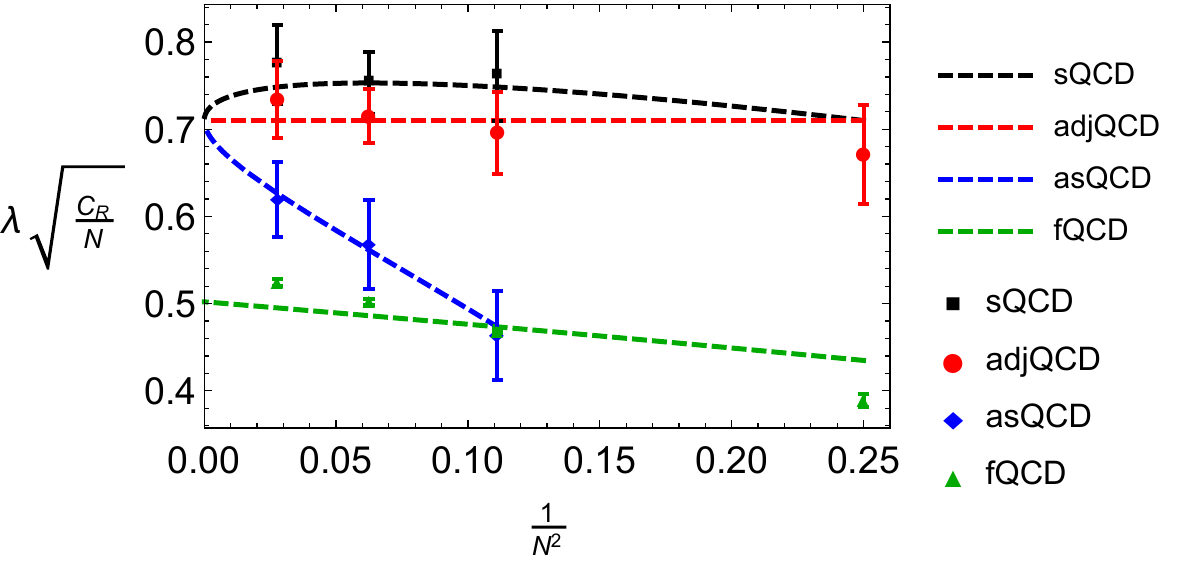}
\caption{Masses of the lightest vector meson versus $N$ in the chiral limit and in various QCD-like theories. Quenched lattice data (points and error bars) are taken from \cite{Lucini:2010kj} for adjQCD, asQCD, sQCD, and from \cite{DelDebbio:2007wk} for fQCD. Lattice data are expressed in units of lattice size and are compared to the model (\ref{model}) with $\lambda=0.71$ (dashed curves).}\label{figres}
\end{figure}

Quenched lattice data are compared to the model (\ref{model}) in Fig.~\ref{figres} for the four QCD-like theories defined in the introduction. The global constant $\lambda$ has been fitted to the value $0.71$ by a least-square method on all the available lattice data. Several observations can be made. First, our model agrees with orientifold equivalence: Meson masses in asQCD, adjQCD and sQCD converge to the same value as $N$ goes to infinity. Then, the meson masses of $N=3$ fQCD and asQCD are equal since the fundamental and conjugate representations have the same color Casimirs in SU(3). The meson masses of $N=2$ adjQCD and sQCD coincide since the two-indices symmetric color representation is the adjoint of SU(2). Finally, the global evolution of the meson masses versus $N$ predicted by our model is quite compatible with the available lattice data.  

The Casimir scaling hypothesis has already been shown to reproduce the lightest glueball masses versus $N$ in \cite{Buisseret:2011bg} as well as the correct $N$-scaling of baryon masses in a constituent approach \cite{Buisseret:2010na,Buisseret:2011aa}. Here, we extend these results by showing that the Casimir scaling of the string tension is able to reproduce the $N$-dependence of the vector meson masses computed in quenched lattice QCD in the chiral limit for various QCD-like theories. 

\begin{figure}[ht]
\includegraphics[width=10cm]{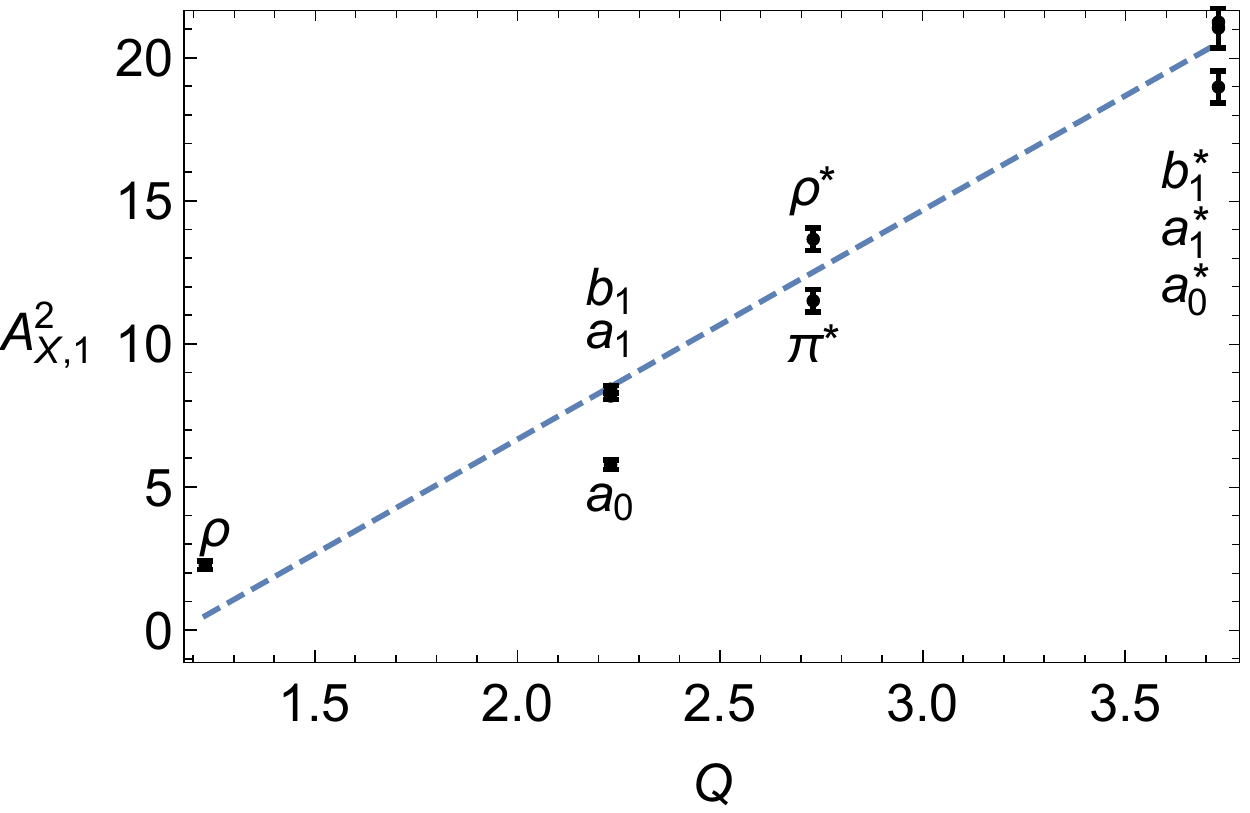}
\caption{Squared large-$N$ scaled masses of several excited mesons in fQCD given by the quantity $A_{X,1}^2$ from the lattice study \cite{Bali:2013kia} versus the global quantum number $Q=1.5\, n+\ell+1.23$ expected from a constituent model (points and error bars). The dashed line is a linear fit, expected from a linear confinement.}\label{figres2}
\end{figure}

The Casimir scaling of the string tension relies on the assumption that meson dynamics is indeed driven by the QCD flux tube, \textit{i.e.} a linear potential in first approximation. It is known that a relativistic quark-antiquark pair with linear potential has a mass spectrum presenting Regge trajectories: the squared meson masses linearly scale with a global quantum number $Q$ which is linear in the radial quantum number $n$ and in the orbital angular momentum $\ell$. More precisely, we have shown that $Q\approx 1.5\, n+\ell+1.23$ in the case of ultrarelativistic quarks with linear potential \cite{Silv09}. In \cite{Bali:2013kia}, scaled masses of some light mesons are computed on a lattice for several values of $Q$, but solely for quarks in the fundamental representation. Masses are fitted as $M_X/\sqrt\sigma=A_{X,1}+A_{X,2}/N^2$ in the chiral limit, and the values of $A_{X,i}$ for various mesons $X$ can be found in Table 24 of \cite{Bali:2013kia}. $A_{X,2}$ contains hyperfine terms that are out of the scope of the present approach: for light mesons the linear confinement is dominant and all spin effects are neglected. When $N\to \infty$, the scaled mass is given by the quantity $A_{X,1}$ and we expect from our model that 
\begin{equation}\label{mesonSR}
\frac{M^2_X}{\sigma}=A_{X,1}^2\propto Q \approx 1.5\, n+\ell+1.23. 
\end{equation}
It is readily seen in Fig.~\ref{figres2} that the trend of the lattice results is compatible with Eq. (\ref{mesonSR}). Similar lattice computations with quarks in higher representations could bring more information about the validity of the Casimir scaling hypothesis for excited mesons but, to our knowledge, such results are not available yet.

The present model only involves a confining interaction. Adding a one-gluon-exchange term would lead, in position space, to a potential of the form $-\frac{C_R}{N}\frac{\alpha_0}{\vert\bm x\vert}$ for a $q\bar q$ pair in color singlet. With the scaling $\bm x \to \bm x/ \sqrt{\sigma_R}$, it is readily checked that this term would lead to a mass spectrum behaving as $M/\sqrt{\sigma_R}=A-B\, C_R/N$, with $A$ and $B$ positive. Moreover $B < A$ since the one-gluon-exchange term is not dominant with respect to confinement for light mesons.  Such a refined model could reproduce spin-dependent differences in the behavior of meson masses versus $N$ computed on the lattice in fQCD \cite{Bali:2013kia}: This is left for future works.

\end{document}